\documentclass[12pt]{article}
\usepackage{epsfig,pstricks}

\newcommand{\nc}{\newcommand}
\nc{\beq}{\begin{equation}}
\nc{\eeq}{\end{equation}}
\nc{\beqa}{\begin{eqnarray}}
\nc{\eeqa}{\end{eqnarray}}

\newwrite\ffile\global\newcount\figno \global\figno=1

\def\writedef#1{}

\input epsf
\def\figin{\epsfcheck\figin}\def\figins{\epsfcheck\figins}
\def\epsfcheck{\ifx\epsfbox\UnDeFiNeD
\message{(NO epsf.tex, FIGURES WILL BE IGNORED)}
\gdef\figin##1{\vskip2in}\gdef\figins##1{\hskip.5in}
\else\message{(FIGURES WILL BE INCLUDED)}%
\gdef\figin##1{##1}\gdef\figins##1{##1}\fi}

\def\figinsert{}
\def\ifig#1#2#3{\xdef#1{fig.~\the\figno}
\writedef{#1\leftbracket fig.\noexpand~\the\figno}%
\figinsert\figin{\centerline{#3}}\medskip\centerline{\vbox{\baselineskip12pt
\advance\hsize by -1truein\center\footnotesize{  Fig.~\the\figno.} #2}}
\bigskip\endinsert\global\advance\figno by1}
\def\endinsert{}

\begin{document}

$\left. \right.$ \vskip 1.5 in
\begin{center}
{\large{\bf Decoupling Dynamical Electroweak Symmetry Breaking} \\
{\it (Talk presented at Beyond the Desert, Castle Ringberg, 1999)} }
\end{center}

\begin{center}
Nick Evans$^1$,

{\small\em {}$^1$Department of Physics,
University of Southampton, Southampton, S017 1BJ, UK.} 
\end{center}

\author{}

\begin{center}October, 1999
\end{center}


\begin{picture}(0,0)(0,0)
\put(350,255){SHEP-99-17}
\end{picture}

\begin{abstract}
The modern precision accelerator data has essentially ruled out the most
obvious model of dynamical electroweak symmetry breaking, technicolour. 
The idea is though well motivated and it is important to construct models
compatible with the precision data of this ilk to motivate experimental
searches. We describe the top-see-saw  and 
flavour universal symmetry breaking 
models that break electroweak symmetry dynamically yet have a decoupling 
limit for all new physics. Limits on the scale of such new physics may be 
placed using precision data and direct search results from the Tevatron.  
\end{abstract}

\newpage

\section{Introduction}

Electroweak symmetry (EWS) is a gauged chiral symmetry of the standard model 
(SM) fermions broken by a mechanism which at this stage remains elusive.
Influenced by the breaking of electro-magnetism in superconductors through
the dynamical formation of an electron pair condensate, and by the breaking 
of chiral symmetry in QCD through a quark  condensate, it is natural
to propose that electroweak symmetry may be broken by a dynamically driven
fermion condensate. If the responsible dynamics were a gauge
interaction then the logarithmic running of the gauge coupling
would naturally provide a separation between the planck and electroweak
scales, {\it ie} a solution to the hierarchy problem. The most obvious such 
extension to the standard model in this vein is technicolour 
\cite{tc} - essentially
a repeat of QCD but with a strong interaction scale of order the weak scale.
Such models though run into problems because in a broken gauge theory there
is a violation of the decoupling theorem \cite{STU}. 
The sector responsible for the
symmetry breaking gives large contributions to parameters in the low energy
theory applicable at LEP and is, at least naively, incompatible. The most 
natural mechanism for the generation of the standard model fermion masses 
in this context is by a feed-down mechanism involving broken gauge 
interactions, extended technicolour (ETC) \cite{etc}. 
ETC also runs into grave trouble - 
in its case accommodating sufficient 
isospin breaking to generate the large top-bottom mass splitting without 
contradicting the precision data \cite{Ttrouble}. 
In this article we discuss recent model
building that overcomes many of the failures of technicolour
\cite{tseesaw,fu}.

Why, given the failures of the arch-type model, and the existence of other
well motivated solutions of the hierarchy problem 
(supersymmetry and large extra dimensions),
should one perserver with dynamical symmetry breaking models? Firstly I
believe that the motivation that inspired technicolour remains even with its 
fall but, most importantly, I do not believe it is the place of theory to 
claim the exclusion of entire paradigms. In the current era of 
particle physics it is the theorist's 
role to provide as wide a variety of viable models for experiment to eventually
differentiate between. 

The dynamical models I discuss below are intended to 
provide insight into how dynamical symmetry breaking might manifest 
in nature and hence inspire experimental searches. The biggest success of
these most recent models is that they are 
compatible with the precision data because they have a decoupling limit in
which low energy predictions are precisely those of the standard model. 

I will begin in Section 1 by reviewing technicolour and the pit falls that
must be avoided. The first example of a dynamical EWS 
breaking model with a decoupling limit 
proposed was top condensation \cite{topc} which I review in section 2 -
the model is though ruled out by the small measured top mass. 
In section 3 I describe
recent models that successfully implement a decoupling limit in dynamical
symmetry breaking models \cite{tseesaw,fu}. 
Finally in section 4 I discuss the experimental
limits on the scale of the new physics proposed in these models both from
precision data and from direct searches at the Tevatron. Much of the work 
reported here was carried out with Gustavo Burdman and Sekhar Chivukula in
\cite{fu, precision, direct}.

\section{Technicolour and its failures}

The simplest model of dynamical EWS breaking is technicolour \cite{tc}. 
We assume there is an $SU(N_{TC})$  
gauge group acting on say a single
electroweak doublet of left handed ``techni-quarks'', $(U,D)_L$, and
two electroweak singlet right handed techniquarks, $U_R$, $D_R$. The 
techniquarks are massless so there is an $SU(2)_L \times SU(2)_R$ chiral 
symmetry. We assume the asymptotically free SU(N) group becomes strong at a 
scale ${\cal O}$(1 TeV), generating techniquark condensates, $\langle \bar{U} 
U \rangle, \langle \bar{D} D \rangle \neq 0$, which break the chiral 
symmetry to the vector subgroup. The chiral EWS is broken. The Goldstones
eaten by the W and Z are the Goldstones of chiral symmetry breaking, the
technipions. The weak scale $v$ is traded for the technipion decay constant,
$F_\pi$. Such a model would be characterized by the discovery of technihadrons
at the TeV or so scale.

This sort of model though gets in trouble with the precision electroweak data
from LEP and SLD \cite{STU}. 
In a broken gauge theory particles with masses violating 
the gauge symmetry do not decouple. These effects enter through oblique
corrections which can be parameterized by the three parameters $S,T,U$
\cite{STU}. The
$S$ parameter turns out to essentially count the number of such massive 
particles. One can estimate the contribution to S from massive, strongly
interacting fermions by scaling up normal QCD data to the appropriate scale
\cite{STU}. 
The result for technifermions is 
\begin{equation}
\Delta S_{TC} \simeq N_{TC} N_{D} ~ 0.1
\end{equation}
where $N_D$ is the number of doublets. The experimental limit on $S$ 
(assuming a heavy higgs) is $-0.27 \pm 0.12$ \cite{data}! Even a very minimal 
one doublet SU(2) technicolour theory appears ruled out. Of course it is
possible that there are other pieces of new physics contributing to S 
with negative sign or
that the naive scaling of QCD data might be inapppropriate to the technicolour
dynamics. In any case at most a relatively minimal technicolour sector seems
possible.

Breaking EWS is not the only job a technicolour model must accomplish. The 
SM fermions must also be given their masses. The usual mechanism considered
is extended technicolour. At high scales the technicolour group is unified
with the flavour symmetries of the SM fermions. This larger symmetry is assumed
to be broken down to technicolour leaving massive gauge bosons which can feed
the technifermion condensate down to provide the SM fermion masses. One
finds
\begin{equation}
m_f ~ \simeq ~ {g^2_{ETC} \over M^2_{ETC}} \langle \bar{T} T \rangle
~ \simeq ~ {g^2_{ETC} \over M^2_{ETC}} 4 \pi F^3_{TC}
\end{equation}
The gauging of the SM flavour symmetries is though a dangerous game and one
may expect to find flavour changing neutral currents in the theory mediated
by single gauge boson exchange. To suppress such contributions to 
$K^0-\bar{K}^0$ mixing requires $M_{ETC} \geq 600$ TeV. Such an ETC gauge
boson can only generate a fermion mass of 0.5 MeV though which is well 
short of the second family quark masses. This is a long standing problem
with ETC.

A second problem is the generation of the large top mass. A 175 GeV 
fermion mass
would require a 1 TeV or so ETC gauge boson. The interactions of this light
gauge boson must violate custodial isospin since the bottom quark is so
much lighter than the top. Including such a isospin violating gauge boson 
in the loops of technifermions generating the W and Z masses gives 
contributions to $\Delta \rho ~(\equiv \alpha T) \simeq 12 \%$ \cite{Ttrouble}
- two orders of magnitude above the experimental limit!

The lessons of technicolour appear to be that there are no extra electroweak
doublets beyond the SM and that 
the SM fermion masses do not result from a simple
feeddown mechanism.

\section{Top condensation and its failure}

Top condensation models \cite{tc}
were a first attempt to avoid the excessive baggage 
of technicolour. Inspired by the large top mass it was suggested that the
top may play a unique role in EWS breaking - perhaps the ``top is the 
technifermion'' and it is a $\langle \bar{t} t \rangle$ condensate that 
breaks EWS. The simplest model is to introduce a four fermion interaction
acting on the top
\begin{equation}
{\cal L} = {\kappa \over M^2} \bar{\psi}_L t_R \bar{t}_R \psi_L 
\end{equation}
where we might imagine some broken gauge theory was providing the origin
of the interaction. At least at large N, the model can be solved and the 
behaviour of the condensate as a function of $\kappa$ is shown in Fig 1.
There is a critical coupling at which chiral symmetry breaking switches 
on. At large $\kappa$ the condensate flattens out to of order the scale 
$M$. If $M \simeq 1$ TeV then arranging $\kappa$ so that the correct 
top mass/EWS breaking scale is realized is relatively easy. If $M \gg 1$
TeV then one must fine tune $\kappa \rightarrow \kappa_c$ to achieve
such a low scale as $v$. Below the scale $M$ the effective theory contains 
a higgs boson which is a bound state of the top quark.
The higgs mass at tree level can be calculated
from resumming top loops in the four top scattering amplitude and at large N
is given by $2 m_t$. One may also estimate the relation between $m_t$ and
$v$ through a loop diagram and here is where the theory runs into trouble.
To generate $v \simeq 250$ GeV requires $m_t \simeq 600$ GeV (assuming 
$M \simeq 5$ TeV)!

\begin{figure}
\epsfxsize 8cm \centerline
{\epsffile{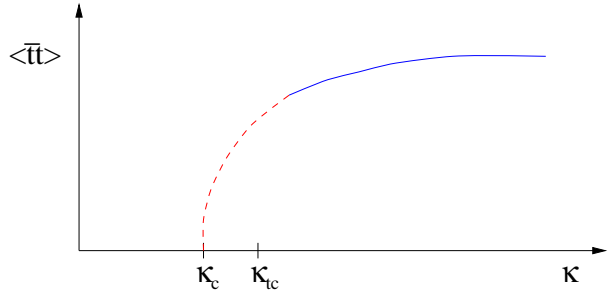}}
\caption{The top condensate as a function of coupling in the four 
fermion interaction theory.
}
\end{figure}

It is possible to combine top condensation and technicolour \cite{tc2}
to lessen the 
$\rho$ parameter problems that technicolour alone suffered. One allows
EWS to be broken by technicolour whilst a direct top condensate supplies
the top mass without a light ETC gauge boson. Such models have most of
the troublesome baggage of technicolour remaining though.

\section{Viable dynamical symmetry breaking models}

We move now to discuss models that are compatible with all low energy data.
The archtype was provide by Dobrescu and Hill in their top see-saw model
\cite{tseesaw}.
Their model provides a mechanism for reconciling $m_t$ with the idea that 
a top condensate provides the entirety of $v$. A similar idea was
proposed in \cite{vector}.

The trick is to use the 
left handed top as a ``technifermion'' but introduce a new field $\chi_R$
to be the right handed technifermion. $\chi_R$ has the same quantum numbers 
as the $t_R$ and is bound into a massive 
Dirac fermion with a partner $\chi_L$  which shares it's quantum numbers
($m_{\chi} \simeq 3$ TeV). 
We now imagine an interaction of the form (3) that is strong and drives a
$\langle \bar{t}_L \chi_R \rangle$ condensate that breaks EWS at the 
scale $v$. A mass of order 600 GeV has been generated between $t_L$ and 
$\chi_R$. To generate the top mass we include a mass term betwen $\chi_L$
and $t_R$ - this is gauge invariant so we would have to explain why it 
wasn't there if it wasn't! The result of all these masses, if we choose
the EWS singlet masses correctly,  is a see-saw
like mass spectrum with a massive eigenstate (1-5 TeV) and a light
eigenstate, the top (175 GeV). It may seem a bit strange that $\chi_R$
which is part of a Dirac fermion with mass of several TeV can participate in 
dynamics that gives rise to a scale of $v$. For this to be possible we require
that the scale $M$ in (3) be larger than $\chi$'s  
Dirac mass term so the dynamics
is really above that scale. The fact that the scale $v$ emerges hints at 
a degree of fine tuning - in this sense it is best if $M$ is not too large.

The higgs in these models is a bound state of $t_L$ and $\chi_R$ and, at
large N, as in the top condensate model, has a mass of twice the EWS breaking
mass, ie 1.2 TeV. This mass is only the tree level mass and does not take
into account the running of the quartic coupling between $M$ and the weak 
scale. This running is quite strong and for large values of $M$ will display
the fixed point behaviour of the SM couplings. We expect for $M < 10$ TeV
that the physical higgs mass will be between 400-600 GeV. 

The appealing aspect of this model is that it has a decoupling limit. 
$\chi$ is an electroweak singlet and so its mass may be taken to infinity
where it will decouple completely from the low energy theory leaving the 
SM as the effective field theory. To maintain the physical top mass
the ratio of the mass between $\chi_L$ and $b_R$ to that of the $\chi_L-\chi_R$
mass must be kept constant in this limit. Of course taking the extreme limit 
of $M \rightarrow \infty$ introduces fine tuning as discussed above but 
if $M \simeq 3+$ TeV the decoupling is almost complete and the fine tuning 
``barely'' present \cite{tseesaw}. 

Extending this type of model to include masses for all the SM fermions
is relatively easy. One example is the flavour universal EWS breaking model
\cite{fu}.
The top mass is no longer a direct measure of $v$ in the see-saw model so
there is in fact no reason to use it, or it alone, to break EWS. In
the flavour universal model all the SM fermions participate equally in
EWS breaking. We introduce two Dirac singlet fermions ($\chi$ and $\omega$)
 with masses of 3 TeV or so and the quantum numbers of
the SM fermion's right handed spinor for each SM fermion. 
A strong interaction 
is assumed to cause condensation between the left handed SM fermion
and its $\chi_R$ field. A mass term is included between $\omega_L$ and
the right handed SM fermion. The SM fermion mass then results from a mass
mixing between the two massive singlets according to graphs such as
\vspace{0.2cm}

\epsfxsize 8cm \centerline
{\epsffile{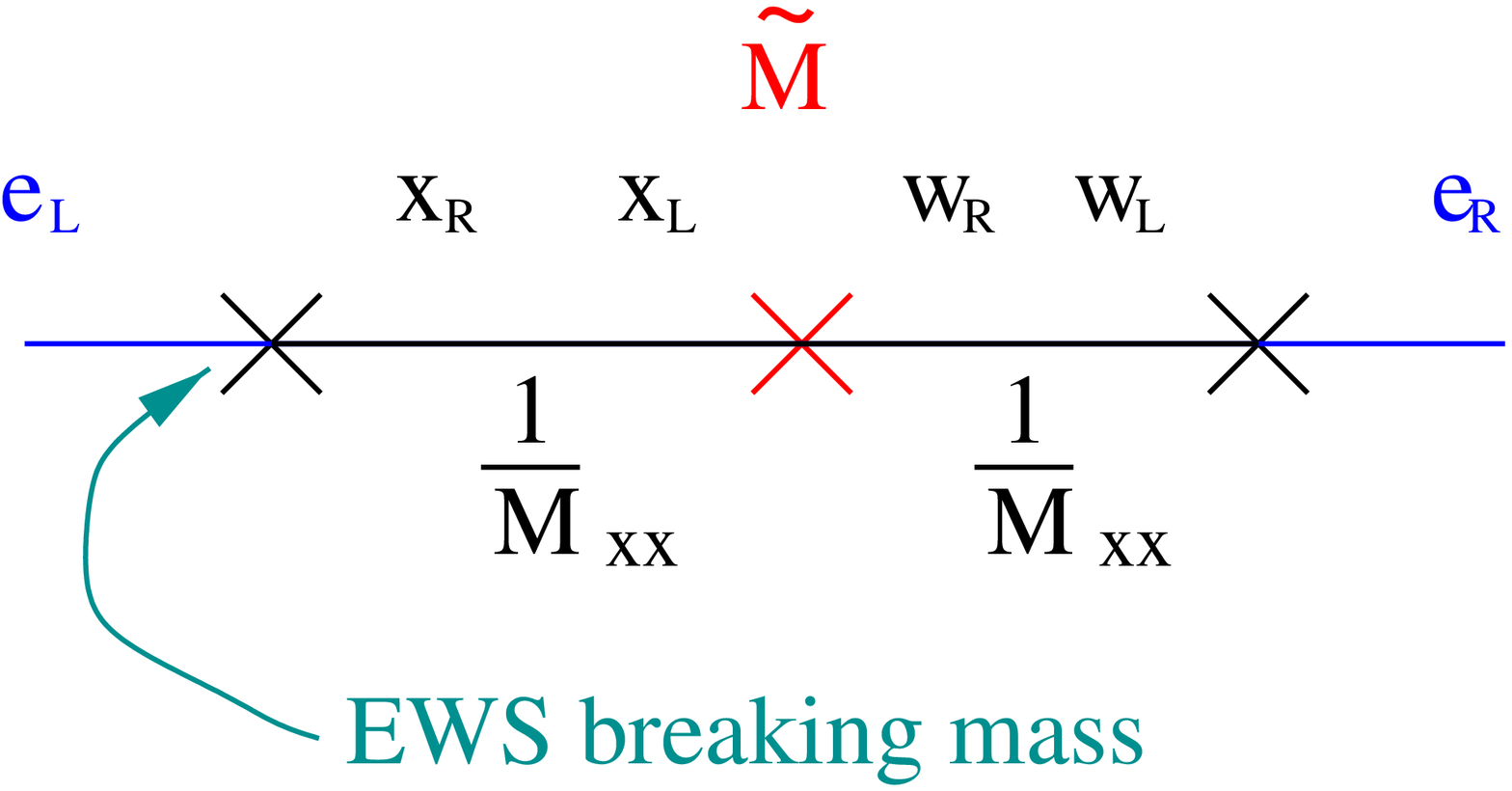}}

The SM fermion masses are simply the result of mass terms, $\tilde{M}$,
chosen in
the singlet sector - the problem of flavour is defered to a higher scale.
This mechanism introduced in \cite{moose} is essentially a way of introducing 
yukawa couplings into dynamical models. Since all the SM fermions
participate equally in EWS breaking the EWS breaking masses between the 
SM fermions and the singlet sector are reduced by a factor of $\sqrt{N_D/3}$
and the higgs mass is approximately 350-450 GeV (with running of the 
quartic coupling the mass could be as low as 300 GeV). 

More complete models of both the top see-saw and the flavour universal 
EWS breaking model exist in the literature. The origins of the strong
coupling are broken, strong, gauged flavour symmetries. For example to
generate a top condensate one must have an interaction that acts solely
on the top - one possibility is to gauge the SU(3) colour group of the
top separately from that of the rest of the standard model
\cite{tseesaw}. At the 3 TeV
or so scale this extended gauge symmetry is broken to the SM leaving a
colour octet of massive strongly interacting gauge bosons. These top colour
interactions are responsible for the top condensation. To distinguish
between the top and bottom quarks these interactions must be chiral. 
The flavour universal models suggest the gauging (and then breaking)
of the chiral family symmetry
groups of the standard model or the full SU(12) flavour symmetry of the 
standard model left handed fermions \cite{fu}. One might worry that as in ETC 
FCNCs will be generated.  In fact if we are careful to preserve the
$SU(3)^5$ chiral flavour symmetry of the standard model, which is responsible
for the SM's GIM mechanism, then the GIM mechanism persists above the 
weak scale and these symmetries can be gauged at scales of only a few TeV
\cite{moose}. 
Since this class of model requires different interactions for the left handed 
doublets from the right handed fermions such a scheme is very natural in this 
context.

\section{Experimental limits}

The dynamical symmetry breaking models described above have been engineered
to have a decoupling limit and hence to avoid making an experimental 
prediction! However, the desire to avoid fine tuning requires that the 
scale of the new physics is actually not too far above the weak scale. It is
therefore possible to place meaningful
lower limits on the scale of the dynamics from
precision EW data and direct search limits from the Tevatron.

The new physics in the models enters the precision data in two ways. Firstly
there is mixing between the SM fermions and EWS singlets which will give
rise to corrections to the SM Z-fermion couplings of the form
\begin{equation}
\delta g_{f} \simeq -{e \over s_\theta c_\theta}
{Q_f s_{\theta_w}^2 } m_{mix}^2 / m_\chi^2~
\end{equation}
where $m_{mix}$ is the mixing mass which we expect of order a few 100 GeV
and $m_\chi$ is the Dirac mass of the singlet. Assuming 
flavour universal mixing and fitting to the data places
a limit of 1.9, 2.6 TeV on $m_\chi$ for $m_{mix} = 100, 200$ GeV.

The flavour gauge bosons may also correct the vertices of the SM fermions
they act on and provide corrections to the $\rho/T$ parameter through
loops of top quarks. In \cite{precision}
we have performed a global fit to the Z-pole
data including these effects. The $95\%$ confidence level limits on the 
mass scale of the new interactions in a variety of models
when their coupling is the critical coupling from the NJL model are
\begin{equation}
\begin{array}{lc}
{\rm Top~ colour} & M(\kappa_c) \geq  1.3 ~{\rm TeV} \\
{\rm Left~ handed~ quark~ family~ symmetry} & M(\kappa_c) \geq 2 ~{\rm TeV} \\
{\rm Left~ handed~ SU(12)~ flavour~ symmetry}   & M(\kappa_c) \geq 2~{\rm TeV}  \\
\end{array}
\end{equation}
These bounds assume a 100 GeV higgs mass but are in fact fairly 
insensitive to the higgs mass since the higgs mass enters only 
logarithmically in the precision variables whilst the mass scale in
these corrections enter quadratically. The precision data currently favours
a low standard model 
higgs mass $m_h \leq 260$ GeV (the bound rises to 400 GeV if the SLD
forward backward asymetry measurement is not included) whilst the models 
we have discussed have a higgs mass in the $300-600$ GeV range. The 
precision limit is though extremely sensitive  to new physics - in
particular positive contributions to the $\delta \rho/T$ parameter such
as are provided by these flavour gauge bosons can restore heavier higgs masses
to agreement with the precision data. For example the top coloron model with
$M(\kappa_c)$ of order 2 TeV is compatible with a 400 GeV higgs.

Direct search limits on the flavour gauge bosons may be obtained from the
Tevatron Run I data. Top colour gives enhanced top production, the 
larger flavour symmetry models enhance $q \bar{q}$ production and can make
use of the bottom quark content of the proton to make single top events. 
Finally the SU(12) flavour model which involves the leptons gives contributions
to Drell-Yan production. These limts are currently under study \cite{direct}
and place bounds of 1-3 TeV on the flavour gauge bosons.
Expectations for Run II's limits are that they will be 
competetive with the precision data and probe scales of 
order 3+ TeV in these dynamical symmetry breaking models.

\section{Conclusions}

The idea that EWS is broken by a dynamically generated fermion
condensate offers the possibility of a natural and low scale extension
of the SM. Technicolour was the obvious first model to propose since it is
simply a repeat of QCD. However, the precision data is incompatible
with an extended symmetry breaking sector. Top condensation
was proposed as a dynamical symmetry breaking model with a minimum of
new physics and provides a natural explanantion for a heavy top. In fact
the top turns out to be too heavy in this scheme. The top see-saw model
resolves this problem by introducing singlet fermions and a see-saw
mass mechanism that gives a lightest mass eignestate that can be 
interpreted as the top quark. The flavour universal symmetry breaking 
model extends the idea to include masses for the full set of SM fermions.
These models have a decoupling limit since all the additional fields
beyond the SM fields are EWS singlets.
The dynamics is assumed to result from broken gauged flavour symmetries. 
Precision data and Tevatron direct searches put a lower limit of order
a few TeV on these models. The Tevatron at Run II has discovery potential.

The models discussed though are not complete models in any sense. The origin
of the SM fermion masses is deferred for example. That the dynamics is 
the result of gauge interactions broken close to their strong scale so they
can nevertheless generate condensation themselves requires unproven 
gauge dynamics \cite{georgi}. 
The hope is that the models provide examples of how dynamical
symmetry breaking might be realized in nature and guidance for
experimental searches. In the end experiment must surely be an essential
guide to the true model of EWS breaking.

\noindent{ \bf Acknowledgements:} NE's work is supported by a PPARC Advanced
Fellowship.

\end{document}